\documentclass[showpacs,twocolumn,prb]{revtex4}
\usepackage{bm}
\usepackage{graphicx}
\usepackage{amsmath}
\usepackage{amssymb}
\usepackage{color}

\begin{document}

\title{Semiclassical theory of the photogalvanic effect in non-centrosymmetric systems}

\author{E.\, Deyo,$^1$ L.\,E.\,Golub,$^2$,  E.\,L. Ivchenko$^2$, B.\,Spivak,$^1$}
\affiliation{$^1$  Department of Physics, University of
Washington, Seattle, WA 98195, USA}
\affiliation{$^2$Ioffe Physico-Technical Institute of the Russian
Academy of Sciences, 194021 St.~Petersburg, Russia}

\begin{abstract}
We develop a semiclassical theory of nonlinear transport and the
photogalvanic effect in non-centrosymmetric media. We show that
terms in semiclassical kinetic equations for electron motion which
are associated with the Berry curvature and side jumps give rise
to a dc current quadratic in the amplitude of the ac electric
field. We demonstrate that the circular photogalvanic effect is
governed by these terms in contrast to the linear photogalvanic
effect and nonlinear I-V characteristics which are governed mainly by the
skew scattering mechanism. In addition, the Berry curvature
contribution to the magnetic-field induced photogalvanic effect is
calculated.
\end{abstract}


\maketitle

\section{Introduction and Phenomenological description}

Recently, considerable progress was made in the semiclassical theory of the anomalous Hall effect
(see the review article, Ref.~\onlinecite{Sinitsyn}). The renewed theory
based on the generalized semiclassical Boltzmann equation removes
the previous controversies surrounded this effect. This theory takes into account the following
mechanisms of the anomalous Hall effect: (i)~the skew scattering
contribution which appears on going beyond the Born approximation
\cite{Smit56,Luttinger58}, (ii)~the contribution to the electron
velocity due to the Berry curvature in the electron Bloch
wavefunction,~\cite{ChangNiu95,Niu99,Haldane} and the contributions of side-jumps,
i.e., coordinate shifts at the scattering events,~\cite{Berger,Sinitsyn06} (iii)~to the
electron velocity and (iv)~to a change of the electron energy upon a
scattering in the presence of an external electric field. In the light of this advance in linear
transport physics, the semiclassical approach to nonlinear
transport should be revised as the next natural step. In the
present paper we develop a semiclassical theory of a \textit{dc}
electric-current response quadratic in an external \textit{dc} or
\textit{ac} electric field. The former response is a nonlinear
contribution to the I-V characteristics while the latter can be
interpreted as a low-frequency photogalvanic effect. The
microscopic theories of photogalvanic effects induced by light
waves of sufficiently high frequencies lying in the terahertz,
infrared or visible regions are based on quantum-mechanical
calculations. They make allowance for both the skew and side-jump
mechanisms (usually called respectively the ballistic and the shift contributions) and have no need for the introduction of the Berry
curvature, for details see the books
Refs.~\onlinecite{SturmanFridkin,Ivchenko_book,GanichevPrettl_book}.
On the other hand, the existing theories of static currents
quadratic in the electric field regard the skew contributions
only, see Refs.~\onlinecite{Bloch1,Bloch2}. The
shift contribution to the I-V characteristics has been considered
in the regime of streaming achieved in strong electric
fields.~\cite{StreamingShift} As for the Berry-curvature mechanism
of  nonlinear transport, to the best of our knowledge, it has
not been analyzed yet.

Phenomenologically, the electric current density ${\bm j}$ which is
quadratic in the electric field is described by
\begin{equation} \label{int1}
j_{\lambda} = \Lambda_{\lambda \mu \nu} E_{\mu} E^*_{\nu} + T_{\lambda \mu \nu \eta}
E_{\mu} E^*_{\nu} q_{\eta}\:,
\end{equation}
where ${\bm \Lambda}$ and ${\bm T}$ are material tensors, ${\bm
E}$ and ${\bm q}$ are the electric-field amplitude and the wave vector
of the {\it ac} electromagnetic wave, $\lambda, \mu, \nu$ and
$\eta$ are the Cartesian coordinates.
In the particular
limit of a static field (the
wave frequency $\omega \to 0$) the amplitude ${\bm E}$
reduces to the {\it dc} electric field, and ${\bm q} \to 0$. The
second term in the right-hand side of Eq.~(\ref{int1}) describes
the photon drag effect. It was first observed and described as the
high-frequency Hall effect by Barlow.~\cite{Barlow} In the
quantum-mechanical description of optical transitions, the drag
effect arises due to momentum transfer from photons to charge
carriers.~\cite{Grinberg,Ryvkin80}
This current can be induced in both centro- and
non-centrosymmetric systems. The semiclassical theory of this
effect is well developed, see, e.g., Refs.~\onlinecite{Gurevich,PerelPinskii_73} and
the review Ref.~\onlinecite{LEGurevich80}, and
here we focus completely on the first term in the right-hand side
of Eq.~(\ref{int1}). It can be conveniently decomposed as
\begin{equation}
\label{int2}
j_{\lambda} = \chi_{\mbox{}_{\lambda \mu \nu}} \{ E_{\mu} E^*_{\nu} \}_{\rm s} + \gamma_{\lambda \mu}\
{\rm i} \left( {\bm E} \times {\bm E}^* \right)_{\mu}\:,
\end{equation}
where $\{ E_{\mu} E^*_{\nu} \}_{\rm s}$ is the symmetrized product
$( E_{\mu} E^*_{\nu} + E^*_{\mu} E_{\nu} )/2$. The two
contributions in Eq.~(\ref{int2}) describe, respectively, the
linear and circular photogalvanic effects, or LPGE and
CPGE.~\cite{SturmanFridkin,Ivchenko_book,GanichevPrettl_book} At $\omega \to 0$, the
vector ${\bm E}$ is certainly real and the CPGE vanishes while the
LPGE reduces to the quadratic conductivity.
Since
the current density is a polar vector and, therefore, odd under
the operation of space inversion, while the square of the electric
field is even, the CPGE and LPGE are allowed only in materials
that lack inversion symmetry, respectively, in gyrotropic media
and piezoelectrics.

We assume that, in addition to the $dc$ or $ac$ electric field, a
static magnetic field ${\bm B}$ is applied to the sample, so that
the coefficients $\chi_{\mbox{}_{\lambda \mu \nu}}$ and
$\gamma_{\lambda \mu}$ are ${\bm B}$-dependent. In the present
work we keep only zero- and first-order terms in ${\bm B}$, in
which case one has
\begin{eqnarray} \label{int3}
\label{chi_B}
\chi_{\mbox{}_{\lambda \mu \nu}}({\bm B}) &=& \chi^{(0)}_{\mbox{}_{\lambda \mu \nu}}
+ \beta^{\rm L}_{\lambda \mu \nu \eta} B_{\eta}\:,
\nonumber \\
\gamma_{\lambda \mu}({\bm B}) &=& \gamma_{\lambda \mu}^{(0)}+\beta_{\lambda \mu \nu}^{\rm C} B_{\nu} \:.
\end{eqnarray}
Since the homogeneous magnetic field is
invariant under space inversion the tensors ${\bm \beta}^{\rm
L}$ and ${\bm \beta}^{\rm C}$ are also nonzero only in
non-centrosymmetric media.

\section{Generalized Boltzmann equation}

Since the photon drag current is neglected we assume the {\it ac}
electric field ${\bm E}(t)$ to be homogeneous and ignore effects
related to the magnetic field of the electromagnetic wave. This
means that the electron distribution function $f_{\bm p} \left( t
\right)$ is independent of the spatial coordinate ${\bm r}$. Since
in this paper we ignore the spin-orbit interaction this function
is also spin-independent. The kinetic equation for $f_{\bm p}
\left( t \right)$ has the standard form
\begin{equation} \label{genboltz1}
\frac{\partial f_{\bm p}}{\partial t} +
\dot{{\bm p}} \cdot \frac{\partial
f_{\bm p}}{\partial {\bm p}} = I_{\bm p}\{f\}\:,
\end{equation}
where $\dot{{\bm p}}$ is the time derivative of the
 momentum of the center of mass of the electron wavepacket,
and $I_{\bm p}\{f\}$ is the collision term. The latter consists of two terms,
$I^{({\rm el})}_{\bm p}\{f\}$ and $I^{({\rm inel})}_{\bm p}\{f\}$,
representing the elastic scattering by static
defects or disorder and inelastic scattering by phonons. For simplicity we
assume the typical elastic-scattering time
$\tau$ to be much shorter than that for inelastic scattering,
$\tau_{\rm inel}$. In this case $I^{({\rm el})}_{\bm p}\{f\}$ and $I^{({\rm inel})}_{\bm p}\{f\}$
describe, respectively, the electron momentum and energy relaxation.
Taking into account the Berry curvature, side-jumps under
electron scattering, and skew scattering, the equations for
$\dot{\bm p}$ and $I_{\bm p}\{f\}$ are modified as compared
to the classical Boltzmann equation. Following Refs.~\onlinecite{Bloch1,Bloch2}
(see also Ref.~\onlinecite{Sinitsyn})
we expand the term $I^{({\rm el})}_{\bm p}\{f\}$ into powers of the scattering potential
$V({\bm r})$ and write it as a sum of the scattering rate found in the lowest, second-order,
approximation
\begin{equation} \label{collision}
I^{({\rm el},2)}_{\bm p}\{f\} = \sum\limits_{{\bm p}'} W^{(2)}_{{\bm p}'{\bm p}}
(f_{{\bm p}'} - f_{\bm p})\:,
\end{equation}
\begin{equation} \label{w2}
W^{(2)}_{{\bm p}'{\bm p}} = \frac{2 \pi}{\hbar} \left<  \left\vert V_{{\bm p}'{\bm p}}
\right\vert^2  \right> \delta[\varepsilon_{p'} - \varepsilon_p - e {\bm E}(t)
\cdot {\bm r}_{{\bm p}'{\bm p}}]\:,
\end{equation}
 and the antisymmetric part of the scattering rate (the skew contribution)
\[
I^{({\rm el},a)}_{\bm p}\{f\} = \sum\limits_{{\bm p}'} W^{(a)}_{{\bm p}'{\bm p}}
(f_{{\bm p}'} - f_{\bm p})
\]
calculated beyond the Born approximation (third- and higher-order
corrections): $W^{(a)}_{{\bm p}'{\bm p}} = - W^{(a)}_{-{\bm p}', -
{\bm p}}$. Here $V_{{\bm p}' {\bm p}}$ is the transition matrix element, $\varepsilon_p$ is the electron energy, and the angle brackets indicate averaging over the
random scattering potential. One can see that the
$\delta$-function in Eq.~(\ref{w2}) contains a change of energy of
the scattered electron under the side-jump ${\bm r}_{{\bm p}'{\bm p}}$ in the presence of an external electric field ${\bm E}$, see Ref.~\onlinecite{Sinitsyn06}.
Note that the
inelastic collision term $I^{({\rm inel})}_{\bm p}\{f\}$ also
contains an asymmetric part. Moreover, in the presence of a
magnetic field the asymmetric scattering is allowed even in the
Born approximation.~\cite{Semicond1984}

\subsection{Zero magnetic field}

In the absence of an external magnetic
field one has $\dot{{\bm p}} = e{\bm E}(t)$, where $e$ is
the electron charge. The electric current density reads
\begin{equation} \label{elcurrent}
{\bm j} = \frac{2 e}{V_d} \sum\limits_{\bm p}
\dot{\bm r}_{\bm p} f_{\bm p}\:.
\end{equation}
Here $V_d$ is the macroscopic volume of the $d$-dimensional
sample, $d=3,2,1$ for a bulk sample, quantum well and quantum
wire, respectively, and the factor 2 takes into account the electron
spin degeneracy. The velocity of the electron wavepacket is given
by
\begin{eqnarray} \label{eqmot1}
&&\dot{\bm r}_{\bm p} = {\bm v}_{\bm p} - \dot{{\bm p}}\times \bm{F}_{\bm p} + \delta \dot{\bm r}_{\bm p}\:,  \\
&&\delta \dot{\bm r}_{\bm p} = \sum_{{\bm p}'}
W_{{\bm p}'{\bm p}}^{(2)} \ {\bm r}_{{\bm p}'{\bm p}}\:, \nonumber
\end{eqnarray}
where ${\bm v}_{\bm p} = \partial \varepsilon_p / \partial {\bm p}$ is the electron
conventional velocity.
 The first two terms in Eq.~(\ref{eqmot1}) were derived in Refs.~\onlinecite{Luttinger58,ChangNiu95,Niu99} while the third term was derived in Refs.~\onlinecite{BelIvchStur,Sinitsyn06} (see also Ref.~\onlinecite{Sinitsyn} for a review).
 The Berry
curvature $\bm{F}_{\bm p}$ is defined by ~\cite{Luttinger58,ChangNiu95,Niu99}
\begin{equation}
{\bm F}_{\bm p} =  \frac{\partial}{\partial {\bm p}} \times {\bm \Omega}({\bm p}), \quad
{\bm \Omega}({\bm p}) = {\rm i} \left< u_{\bm k} \Bigg| \frac{\partial
u_{\bm k}}{\partial {\bm k}} \right>
\end{equation}
with ${\bm k} = {\bm p}/\hbar$, and $u_{\bm k}(\bm r)$ being the periodic amplitude of the Bloch
wavefunction. The spatial shift of an electron under the
scattering ${\bm p} \to {\bm p}'$ has the
form~\cite{Sinitsyn06,BelIvchStur}
\begin{equation} \label{sp_shift}
{\bm r}_{{\bm p}' {\bm p}} =  - \frac{ {\rm Im} \left< V_{{\bm p}' {\bm p}}^*
\hat{D}_{{\bm p}' {\bm p}} V_{{\bm p}' {\bm p}}
\right>}{\left< \left\vert V_{{\bm p}'{\bm p}}
\right\vert^2 \right>} + {\bm \Omega}({\bm p}') - {\bm \Omega}({\bm p})\:,
\end{equation}
where
\[
\hat{D}_{{\bm p}' {\bm p}} = \hbar \left( \frac{\partial}{\partial {\bm p}'} + \frac{\partial}{\partial {\bm p}} \right) \:.
\]

\subsection{Nonzero static magnetic field}
The equation for $\dot{\bm p}$ is modified in a magnetic field ${\bm B}$
into~\cite{Sinitsyn,Niu99}
\begin{equation} \label{eqmot2}
\dot{{\bm p}} = e \left[ {\bm E}(t) + \frac{\dot{\bm r}_{\bm p}}{c} \times {\bm B} \right]\:,
\end{equation}
where $\dot{\bm r}_{\bm p}$ is defined by Eq.~(\ref{eqmot1}).

Taking into account the Berry-curvature-induced renormalization of
phase-space volume\cite{Xiao06} the equations for the electron and
electric current densities, $N_d$ and ${\bm j}$, are also
modified as follows
\begin{eqnarray} \label{elcurrent2}
N_d &=& \frac{2}{V_d} \sum\limits_{\bm p} \left( 1 - \frac{e}{c} {\bm B}
\cdot {\bm F}_{\bm p}\right)
f_{\bm p} \:, \nonumber \\ {\bm j} &=& \frac{2 e}{V_d} \sum\limits_{\bm p}
\left( 1 - \frac{e }{c} {\bm B} \cdot {\bm F}_{\bm p} \right)
\dot{\bm r}_{\bm p} f_{\bm p}\:.
\end{eqnarray}
Due to the same reason, in Eq.~(\ref{collision}) one should perform
the replacement\cite{Xiao05}
\begin{equation} \label{replace}
\sum\limits_{{\bm p}'} \rightarrow \sum\limits_{{\bm p}'}
\left( 1 - \frac{e}{c} {\bm B} \cdot {\bm F}_{{\bm p}'} \right)\:.
\end{equation}
Below we take into account contributions to the current
(\ref{elcurrent}) up to the first order in ${\bm F}_{\bm p}$ and
${\bm r}_{{\bm p}'{\bm p}}$. This allows one to approximate the
$\delta$-function in Eq.~(\ref{w2}) by
\begin{eqnarray} \label{deltaerE}
&&\delta[ \varepsilon_{p'} - \varepsilon_p - e {\bm E}(t) \cdot
{\bm r}_{ {\bm p}' {\bm p} }] \\ &&\mbox{} \hspace{1 cm}=
\delta(\varepsilon_{p'} - \varepsilon_p) +
e {\bm E}(t) \cdot {\bm r}_{{\bm p}'{\bm p}}\ \frac{\partial}{\partial \varepsilon_p}
\delta(\varepsilon_{p'} - \varepsilon_p)\:. \nonumber
\end{eqnarray}
Expanding $\dot{\bm r}_{\bm p}$ and $\dot{\bm p}$ up to the first
order in ${\bm F}_{\bm p}$ we can present the solution of the coupled
equations (\ref{eqmot1}) and (\ref{eqmot2}) in the following form
\begin{widetext}
\begin{eqnarray}
\label{dotrdotk}
\dot{\bm r}_{\bm p} &=& {\bm v}_{\bm p} \left( 1 + \frac{e}{c}{\bm B}
\cdot {\bm F}_{\bm p}  \right) - e {\bm E}(t) \times {\bm F}_{\bm p}
- \frac{e}{c} \left( {\bm v}_{\bm p} \cdot {\bm F}_{\bm p} \right) {\bm B} + \delta \dot{\bm r}_{\bm p}\ , \\
\dot{\bm p} &=& e\left[  {\bm E}(t) \left( 1 + \frac{e}{c} {\bm B}
\cdot {\bm F}_{\bm p} \right) + \frac{ {\bm v}_{\bm p} + \delta \dot{\bm r}_{\bm p}}{c} \times
{\bm B} - \frac{e}{c} \biggl({\bm E}(t) \cdot {\bm B} \biggr) {\bm F}_{\bm p} \right]\ . \nonumber
\end{eqnarray}
\end{widetext}

\section{Nonlinear transport in the absence of magnetic field}
The electron distribution function $f_{\bm p}$ is expanded in
powers of the electric-field amplitude ${\bm E}$ defined by
\begin{equation} \label{acfield}
{\bm E}(t) = {\bm E} {\rm e}^{- {\rm i} \omega t} +
{\bm E}^* {\rm e}^{{\rm i} \omega t}\:.
\end{equation}
For the classical Boltzmann equation in the relaxation-time and
effective-temperature approximations one has
\begin{equation} \label{f1}
f^{(1)}_{\bm p}(t) = {\rm e}^{- {\rm i} \omega t} f^{(1)}_{{\bm p} \omega}  + {\rm c.c.} \:,
\quad f^{(1)}_{{\bm p} \omega} = - e \tau_{\omega}
({\bm E} \cdot {\bm v}_{\bm p}) f_0'
\end{equation}
for the linear correction, and
\begin{eqnarray} \label{f2}
&&f^{(2)}_{\bm p} = f_0(\varepsilon_p, \Theta) - f_0(\varepsilon_p)
+ f^{(2)}_2({\bm p})\:, \\
&&f^{(2)}_2({\bm p}) = 2 e^2 \tau {\rm Re}(\tau_\omega) f_0^{\prime\prime} (v_{{\bm p}\mu} v_{{\bm p}\nu} - \overline{v_{{\bm p}\mu}^2} \delta_{\mu\nu}) \{ E_{\mu} E^*_{\nu} \}_s
\:, \nonumber
\end{eqnarray}
for the time-independent second-order correction. Here
the momentum relaxation time $\tau$ is defined by
\begin{equation} \label{tau}
I^{({\rm el},2)}_{\bm p}\{f\} = - \frac{f_{\bm p} - \overline{f_{\bm p}}}{\tau}\:,
\end{equation}
where the bar means averaging over the directions of ${\bm p}$,
hereafter for simplicity we neglect the dependence of
$\tau$ on the electron energy $\varepsilon_p$,
$f_0(\varepsilon_p, \Theta)$ is the Fermi-Dirac distribution
function at the effective temperature $\Theta$ different from the
bath temperature $T$ due to the heating of the electron gas,
$f_0(\varepsilon_p) \equiv f_0(\varepsilon_p,T)$,
a prime means derivative over $\varepsilon_p$: $f_0^{\prime} = \partial f_0(\varepsilon_p)/\partial \varepsilon_p$, and
\[
\tau_{\omega} = \frac{\tau}{1 - {\rm i} \omega \tau}\:.
\]

In the following subsections we will take into account skew-scattering,
Berry-curvature, and side-jump effects, find
corrections to the electron distribution function given by the sum
of (\ref{f1}) and (\ref{f2}), substitute these corrections into
Eq.~(\ref{elcurrent}) and find the electric current proportional
to bilinear products $E_{\mu} E_{\nu}^*$ in accordance with the
phenomenological equation (\ref{int2}). It should be mentioned
that, for an {\it ac} electric field, the current (\ref{int2}) or
(\ref{elcurrent}) is defined as the time average. Taking into
account the definition (\ref{acfield}) of the field amplitude the
tensor components $\chi_{\mbox{}_{\lambda \mu \nu}}(\omega)$ in
Eq.~(\ref{int2}) found in the limit $\omega \to +0$ and those for
the static electric field differ by a factor of 2.\\

\subsection{Skew-scattering contribution}
In this subsection we set all vectors ${\bm r}_{{\bm p}' {\bm p}}$
and ${\bm F}_{\bm p}$ to zero so that the nonlinear current
appears only due to the asymmetrical parts of elastic and
inelastic scattering rates, respectively, $I^{ ({\rm el},a) }_{\bm p} \{ f \}$ and $I^{ ({\rm inel},a) }_{\bm p}\{ f \}$ and one has
\begin{equation}
\label{skew}
{\bm j} = \frac{2 e\tau}{V_d} \sum\limits_{\bm p}  {\bm v}_{\bm p} \left[
I^{ ({\rm el},a) }_{\bm p} \left\{ f^{(2)}_2 \right\} +
I^{ ({\rm inel},a) }_{\bm p}\{ f_0(\varepsilon_p, \Theta) \} \right]\:.
\end{equation}
The second term contributes to the current only in systems
possessing a polar axis.

Equation~\eqref{skew} demonstrates that the skew-scattering processes give rise only to LPGE. The current proportional to the degree of circular polarization of the electric field does not appear due to skew scattering.

\subsection{Berry-curvature related current}
Let us now ignore both the asymmetry of collision rate and
side-jumps at the scattering and analyze the effect of the Berry
curvature on the nonlinear current. Note that in centrosymmetric
systems ${\bm F}_{\bm p}={\bm F}_{ -{\bm p}}$, while in systems with time
reversal symmetry  ${\bm F}_{\bm p}= - {\bm F}_{ -{\bm p}}$.~\cite{Haldane} Thus if the system is both time- and centrosymmetric the Berry curvature vanishes, ${\bm F}_{\bm p}
\equiv 0$.

The nonlinear current is obtained by taking into account
linear-in-${\bm E}$ corrections to the velocity $\dot{\bm r}_{\bm p}$ and to the distribution function, see Eqs.~(\ref{dotrdotk}) and
(\ref{f1}), and time averaging  the sum
\[
{\bm j} = \frac{2e^2}{V_d} \sum\limits_{\bm p}  {\bm F}_{\bm p} \times
{\bm E}(t) \: f^{(1)}_{\bm p}(t)\:.
\]
The result written in terms of the tensors
$\chi^{(0)}_{\mbox{}_{\lambda \mu \nu}}$ and $\gamma_{\lambda
\mu}^{(0)}$ reads
\begin{eqnarray} \label{chi1}
\chi_{\lambda \mu \nu}^{(0)} &=& \frac{2 e^3}{V_d} \, {\rm Re} \left(\tau_\omega\right)  \sum\limits_{\bm p} f_0^{\prime} (\epsilon_{\lambda \mu \eta}
C_{\nu \eta} + \epsilon_{\lambda \nu \eta} C_{\mu \eta})
\:, \nonumber \\
\gamma_{\lambda \mu}^{(0)} &=& \frac{2 e^3}{V_d} \, {\rm Im} \left(\tau_\omega\right) \sum_{\bm p} f_0^{\prime} C_{\lambda \mu}
\:,
\end{eqnarray}
where $\epsilon_{\lambda \mu \eta}$ is
the antisymmetric unit third-rank tensor and
$$C_{\lambda \mu}(\varepsilon_p) = \overline{v_{\bm p \lambda} \: F_{\bm p, \mu}}.
$$
While deriving the equation for $\gamma_{\lambda \mu}^{(0)}$ we
took into account that the sum $\sum_{\eta} C_{\eta \eta}$ makes
no contribution to the current because ${\bm \nabla}_{\bm p} \cdot
[{\bm \nabla}_{\bm p} \times {\bm \Omega}({\bm p})] \equiv 0$.

In the particular case of
$d$-dimensional
degenerate electron
gas with a parabolic dispersion $\varepsilon_p = p^2/2 m^*$ ($m^*$ is the electron effective mass) where $k_{\rm B} T \ll E_{\rm F}$ and $f_0^{\prime} \approx -
\delta(\varepsilon_p - E_{\rm F})$ with $E_{\rm F}$ being the Fermi
energy, Eqs.~(\ref{chi1}) are reduced to
\begin{eqnarray} \label{chidegen}
&&\chi_{\lambda \mu \nu}^{(0)} = - \frac{e m^* d}{2 E_{\rm F}} (\epsilon_{\lambda \mu \eta} C_{\nu \eta} +
\epsilon_{\lambda \nu \eta} C_{\mu \eta}) {\rm Re} \left(\sigma_\omega\right)\:, \nonumber \\
&&\gamma_{\lambda \mu}^{(0)} = -\frac{e m^* d}{2 E_{\rm F}} C_{\lambda \mu}
{\rm Im} \left( \sigma_\omega \right)\:,
\end{eqnarray}
where $\sigma_\omega$ is the linear Drude conductivity,
\[
\sigma_\omega = \frac{N_{d} \, e^2 \tau_\omega}{m^*}\:,
\]
and the value of $C_{\lambda \mu}(\varepsilon)$ is taken at the Fermi energy.

Symmetry analysis allows one to find linearly independent
coefficients in the expansion of the Berry curvature ${\bm F}_{\bm
p}$ in powers of the momentum ${\bm p}$. To illustrate, below we
give examples of such analysis for the bulk crystal symmetries
T$_d$ (zinc-blende lattice) and C$_{6v}$ (wurtzite) and the point
symmetry C$_s$ of a quantum-well structure grown along the
low-symmetry axis $[hhl]$ different from [001], [111] or [110].

In the $T_d$ point group the components of the polar vector ${\bm
v}_{\bm p} \propto {\bm p}$ and axial vector ${\bm F}_{\bm p}$
transform according nonequivalent irreducible representations
F$_2$ and F$_1$. Since the direct product of the representations
F$_2$ and F$_1$ does not contain the identity representation A$_1$
the average of the product $v_{{\bm p} \lambda} F_{{\bm p}, \mu}$
over the directions of ${\bm p}$ vanishes, and the Berry-curvature
related current is forbidden for the $T_d$ symmetry. The zero value of $\gamma_{\lambda \mu}^{(0)}$  agrees
with the fact that the $T_d$ crystal class is not gyrotropic and
forbids the circular PGE.

In crystals of the C$_{6v}$ symmetry with $z$ being the polar axis
the components $v_{{\bm p} x}, v_{{\bm p} y}$ and $- F_{{\bm p},
y}, F_{{\bm p}, x}$ transform according to the representation
E$_2$ and the nonzero averages $C_{\lambda \mu}$ are $C_{xy} = -
C_{yx}$. It follows then that nonzero components of the
photogalvanic tensors are $\chi^{(0)}_{xzx} = \chi^{(0)}_{xxz} =
\chi^{(0)}_{yzy} = \chi^{(0)}_{yyz}$ and $\gamma_{xy}^{(0)} = -
\gamma_{yx}^{(0)}$, and they can be expressed via
$C_{xy}(\varepsilon_p)$ by Eqs.~\eqref{chi1}.
Expanding  ${\bm F}_{\bm p}$  to first order in ${\bm p}$  we have
\begin{equation} \label{polar}
{\bm F}_{\bm p} = A \: \hat{\bm c} \times {\bm p}\:,
\end{equation}
where $\hat{\bm c}$ is a unit vector along the polar axis $z$ and
$A$ is a constant. Note that a similar equation follows for the
continuous group C$_{\infty v}$. From Eq.~\eqref{polar} we obtain
\[
C_{xy}(\varepsilon_p) = \frac{2}{3} A \varepsilon_p,
\]
and using Eqs.~(\ref{chi1}) we get for
arbitrary degeneracy of the electron gas
\begin{equation} \label{chigammacxy2}
\chi^{(0)}_{xzx} = A \ {e m^*} \ {\rm Re}
\left(\sigma_\omega\right) \:,
\quad
\gamma_{xy}^{(0)} = - A \ {e m^*} \ {\rm Im}
\left(\sigma_\omega\right) \:.
\end{equation}

In the band-structure model of a wurtzite-type semiconductor
including the conduction band $\Gamma_{1c}$, and the valence bands
$\Gamma_{6v}, \Gamma_{1v}$ we obtain the following estimation for
the constant in Eq.~(\ref{polar})
\begin{equation} \label{A_polar}
A = \frac{4 \hbar P_{\perp} P_{\parallel} Q}{E_g \Delta_c (E_g + \Delta_c)}\:.
\end{equation}
Here
\begin{eqnarray}
P_{\perp} =  {-{\rm i}  \over m_0} \langle S |p_x| X \rangle =
{-{\rm i}  \over m_0} \langle S |p_y| Y \rangle\:,\:
P_{\parallel} = {-{\rm i} \over m_0} \langle S |p_z| Z \rangle
\:, \nonumber \\
Q = {-{\rm i} \over m_0} \langle Z | p_x | X \rangle = {-{\rm i} \over m_0} \langle Z | p_y | Y \rangle\:, \hspace{2 cm} \mbox{} \nonumber
\end{eqnarray}
$m_0$ is the free electron mass, $p_x = - {\rm i} \hbar \partial/
\partial x$, $E_g$ is the fundamental energy gap, $\Delta_c$ is
the crystal splitting, and $S$, $(X,Y)$ and $Z$ are the
$\Gamma$-point Bloch functions in the $\Gamma_{1c}$, $\Gamma_{6v}$
and $\Gamma_{1v}$ bands, respectively. Note that the polar crystal
symmetry is fixed by a nonzero matrix element $Q$.

In a two-dimensional system of the C$_s$ symmetry with the
interface plane normal to $z$ and the mirror reflection plane
$(yz)$ a single nonzero component of ${\bm F}$ is $F_{{\bm p},
z}$. The first-order term is given by
\begin{equation} \label{Cs}
F_{{\bm p}, z} = A_s \: p_x\:,
\end{equation}
where $A_s$ is a constant. As a result one has
\[
C_{\lambda \nu}(\varepsilon_p) = A_s \varepsilon_p
\delta_{\lambda x} \delta_{\nu z}
\]
and one can readily use Eqs.~\eqref{chi1} to relate the photogalvanic tensors to the real
and imaginary parts of the two-dimensional conductivity
$\sigma_\omega$.

\subsection{Shift currents}

In the semiclassical theory, the shift current consists of two
contributions, the first one is due to the scattering-induced
correction to the electron velocity, see Eq.~\eqref{eqmot1}, and has the form
\begin{equation} \label{jvel}
{\bm j}^{({\rm vl})} = {2 e \over V_d}
\sum_{\bm p} \delta \dot{\bm r}_{\bm p}
\: f_2^{(2)}({\bm p})\:,
\end{equation}
while the other is related to the field-induced correction to the
collision term and can be presented by the time average of
\[
{\bm j}^{({\rm sc})} = {2 e \over V_d} \sum\limits_{\bm p} {\bm v}_{\bm p} \: \delta f_{\bm p}\:,
\]
where
\begin{eqnarray}
\delta f_{\bm p} =&& - 2 e \tau \sum\limits_{{\bm p}'} [f^{(1)}_{\bm p}(t) - f^{(1)}_{{\bm p}'}(t)]
\biggl({\bm E}(t) \cdot {\bm r}_{{\bm p}'{\bm p}} \biggr) \nonumber\\
&&\times  \frac{2 \pi}{\hbar} \left< \left\vert V_{{\bm p}'{\bm p}}
\right\vert^2 \right> \frac{\partial}{\partial \varepsilon_p}
\delta(\varepsilon_p - \varepsilon_{p'})\:.
\end{eqnarray}
Taking into account the identity
$$
{\bm v}_{\bm p} \frac{\partial S(\varepsilon_p)}{\partial \varepsilon_p} =
\frac{\partial S(\varepsilon_p)}{\partial {\bm p}}
$$
the second contribution may be reduced into
\begin{eqnarray}
\mbox{}&& \hspace{1.5 cm}{\bm j}^{({\rm sc})} = \frac{4 \pi e^2\tau}{\hbar V_d^2} \sum\limits_{{\bm p} {\bm p}'}
\delta(\varepsilon_p - \varepsilon_{p'})  \\
&&\times \frac{\partial}{\partial {\bm p}} \left[ (f^{(1)}_{{\bm p}
\omega} - f^{(1)}_{{\bm p}' \omega}) \left< \left\vert V_{{\bm p}'{\bm p}} \right\vert^2 \right> ({\bm E}^* \cdot {\bm r}_{{\bm p}'{\bm p}} ) \right] + {\rm c.c.} \nonumber\:.
\end{eqnarray}

We ignore the shift-induced corrections to antisymmetric part of the scattering rate $W^{(a)}_{{\bm p}'{\bm p}}$ because the combined effect of the Berry curvature and side-jumps is negligible.

In order to obtain simple analytical equations for the
photogalvanic tensors, we simplify the model and assume that the
scattering matrix element $V_{{\bm p}'{\bm p}}$ is the sum of a
main term $V_0$ independent of the electron momenta, and an
additional term $\delta V_{{\bm p}'{\bm p}}$ dependent on the momenta, governing the spatial shift (\ref{sp_shift}) but making
no influence on the relaxation time $\tau$. In this model the
current ${\bm j}^{({\rm sc})}$ can be written as
\begin{eqnarray} \label{jscat}
\mbox{}&& \hspace{1.5 cm}{\bm j}^{({\rm sc})} = - \frac{2 e^3}{V_d^2}  \tau \tau_{\omega}
\sum\limits_{{\bm p} {\bm p}'}
W^{(2)}_{{\bm p}'{\bm p}}\\
&&\times \frac{\partial}{\partial {\bm p}} \left\{ \left[
{\bm E} \cdot \left( \frac{\partial f_0(\varepsilon_p) }{ \partial
{\bm p} } - \frac{\partial f_0(\varepsilon_{p'}) }{ \partial
{\bm p}' } \right)\right] ({\bm E}^*
\cdot {\bm r}_{{\bm p}'{\bm p}} )  \right\} + {\rm c.c.}
\nonumber
\end{eqnarray}
The further simplifications become possible if we expand the shift
(\ref{sp_shift}) in powers of ${\bm p}',{\bm p}$ and retain
quadratic terms. The latter are the lowest nonvanishing terms
taking into account the following properties of the polar vector
${\bm r}_{{\bm p}'{\bm p}}$: spatial shifts under electron
scattering are symmetrical functions of the momenta,
${\bm r}_{-{\bm p}', -{\bm p}} = {\bm r}_{{\bm p}' {\bm p}}$, and
reverse under the interchange of ${\bm p}'$ and ${\bm p}$,
${\bm r}_{{\bm p}, {\bm p}'} = - {\bm r}_{{\bm p}' {\bm p}}$. For bulk
T$_d$- and C$_{6v}$-symmetry crystals the second-order terms are
\begin{eqnarray}
\label{TdC6v}
{\bm r}_{{\bm p}'{\bm p}} &=& \eta (p_z {\bm p} - p'_z {\bm p}')
\hspace{15 mm} ({\rm C}_{6v})
\:, \\
r_{{\bm p}' {\bm p}; x}
&=& \eta_d (p_y p_z - p'_y p'_z)
\hspace{12 mm}({\rm T}_d), \nonumber
\end{eqnarray}
and two other components are obtained by cyclic permutation.
For simplicity, mixed terms of type $p_z p'_{\lambda} -
p_{\lambda} p'_z$ allowed by the C$_{6v}$ point group are
neglected.

For a quantum well of the C$_s$ symmetry, the elementary spatial
shift has the form
\begin{eqnarray}\label{r_Cs}
{\bm r}_{{\bm p}'{\bm p}}  &=& \left[\eta_1 \left(
p_x^2 - {p'_x}^2 \right) + \eta_2 \left(
p_y^2 - {p'_y}^2 \right) \right] \hat{\bm y} \nonumber\\
\mbox{} \hspace{17 mm} && + \hspace{3 mm} \eta_3 \left(
p_x p_y - p'_x p'_y \right) \hat{\bm x} \ ,
\end{eqnarray}
where $\hat{\bm x}$ and $\hat{\bm y}$ are the unit vectors along
the $Ox$ and $Oy$ axes, respectively. By using Eqs.~(\ref{jvel}),
(\ref{jscat}) and (\ref{r_Cs}) we obtain the following nonzero
tensor components for a heterostructure of the C$_s$ symmetry
\begin{eqnarray} \label{velC_s}
\chi^{(0)}_{yxx}({\rm vl}) &=& - \chi^{(0)}_{yyy}({\rm vl})  = 2 (\eta_1 - \eta_2)\, e m^*
{\rm Re}(\sigma_{\omega})\ ,  \nonumber \\
\chi^{(0)}_{xxy}({\rm vl}) &=&   \eta_3 {e m^*} {\rm Re}(\sigma_{\omega})\: , \nonumber \\
\chi^{(0)}_{yyy}({\rm sc}) &=& 2 \chi^{(0)}_{xxy}({\rm sc})  = {2} (\eta_1 + \eta_2)\, e m^*
{\rm Re}(\sigma_{\omega})\ , \nonumber \\
\gamma^{(0)}_{xz}({\rm sc}) &=& (\eta_1 + \eta_2) \, {e m^*} {\rm Im}(\sigma_{\omega})\ .
\end{eqnarray}
Since the tensor $\chi^{(0)}_{\lambda \mu \nu}$ is symmetrical
with respect to interchange of the second and third indices the
components $\chi^{(0)}_{xyx}({\rm vl})$ and $\chi^{(0)}_{xyx}({\rm
sc})$ are also nonzero.

\section{Nonlinear transport in the presence of magnetic field}

Here we outline the derivation of the Berry-curvature related
photocurrent linear in a static magnetic field $\bm{B}$. In the
presence of a magnetic field the kinetic equation has the form
\begin{equation}
\frac{\partial f_{\bm p}}{\partial t} + \dot{\bm p} \cdot {\partial f_{\bm p} \over \partial {\bm p}}
= \sum_{{\bm p}'} \left( 1 - \frac{e }{c} {\bm B}
\cdot {\bm F}_{{\bm p}'} \right) W^{(2)}_{{\bm p}'{\bm p}}
(f_{{\bm p}'} - f_{\bm p})\:,
\end{equation}
where $\dot{\bm p}$ is given by Eq.~(\ref{dotrdotk}). It is
instructive to introduce the auxiliary function
\[
\phi_{\bm p} = \left( 1 - \frac{e}{c} {\bm B}
\cdot {\bm F}_{\bm p} \right) f_{\bm p}\ .
\]
Then in the linear-in-$\bm B$ approximation
we have the following kinetic equation for this function:
\begin{eqnarray}
\label{auxiliary}
\frac{\partial \phi_{\bm p}}{\partial t} &+& \frac{e^2}{c} {\bm E} \cdot {\partial \over \partial {\bm p}}
[({\bm B} \cdot {\bm F}_{\bm p}) \phi_{\bm p} ]\\
&+& e \ \left[ {\bm E} + \frac{{\bm v}_{\bm p}}{c} \times {\bm B} -
\frac{e}{c} ({\bm E} \cdot {\bm B}){\bm F}_{\bm p} \right] \cdot
{\partial \phi_{\bm p} \over \partial {\bm p}} \nonumber \\
&=& - \frac{1}{\tau} \left[ \phi_{\bm p} - \overline{\phi_{\bm p}}
+ \frac{e}{c} ({\bm B} \cdot {\bm F}_{\bm p})\overline{\phi_{\bm p}} \right]\ .
\nonumber
\end{eqnarray}
Let us introduce the notations for corrections $\phi_{\bm p}^{(E^mB^n)}$ to the auxiliary function with $m = 0,1,2$; $n = 0,1$, e.g., $\phi_{\bm p}^{(E)}$, $\phi_{\bm p}^{(B)}$, $\phi_{\bm
k}^{(EB)}$ etc.
From Eq.~(\ref{elcurrent2}) we find that the following expression for
the electric current quadratic-in-$\bm E$ and linear-in-$\bm B$
\begin{eqnarray}
\label{j}
&&{\bm j}^{(E^2B)} = {2e \over V_d} \sum_{\bm p} \left[
{\bm v}_{\bm p} \phi_{\bm p}^{(E^2B)} \right. \\ &&\left.
- e ({\bm E} \times {\bm F}_{\bm p}) \phi_{\bm p}^{(EB)}
+ \frac{e}{c} {\bm F}_{\bm p} \times \left( {\bm v}_{\bm p} \times{\bm B} \right) \phi_{\bm p}^{(E^2)}
\right] \ . \nonumber
\end{eqnarray}

As an example we further consider a 2D system of C$_s$
symmetry and take ${\bm F}_{\bm p}$ in the form ${\bm F}_{\bm p}
\parallel z$, $F_{{\bm p},z} = A_s p_x$, see Eq.~(\ref{Cs}).
Solving the kinetic equation \eqref{auxiliary} one can obtain the linear- and quadratic-in-$\bm E$ corrections
at $B = 0$ given by Eqs.~(\ref{f1}) and (\ref{f2}) and linear-in-$\bm B$
corrections in the form
\begin{equation}
\label{phi_B}
\phi_{\bm p}^{(B)} = -\frac{e}{c} B_z F_{{\bm p}, z} f_0(\varepsilon_p) \ ,
\end{equation}
\begin{equation}
\phi_{\bm p}^{(EB)} = {(e \tau_\omega)^2 \over mc } f_0' {\bm v}_{\bm p} \cdot
({\bm B} \times{\bm E}) {\rm e}^{- {\rm i}\omega t} + {\rm c.c.}\ ,
\end{equation}
and, finally,
\begin{equation}
\phi_{\bm p}^{(E^2B)} = \frac{e^3}{c} \tau \tau_{\omega} B_z {\bm E}^* \cdot {\partial \over \partial {\bm p}}
[  F_{{\bm p}, z} ({\bm E} \cdot {\bm v}_{\bm p}) f_0' ] + {\rm c.c.}
\end{equation}
In the above expressions, we neglected terms which do not contribute to the current.
For the particular case of C$_s$ symmetry
nonzero components of the magnetophotogalvanic tensors $\beta^{\rm
L}_{\lambda \mu \nu \eta}$ and $\beta_{\lambda \mu \nu}^{\rm C}$
defined according to Eq.~(\ref{chi_B}) become
\begin{subequations}
\begin{eqnarray}
\beta^{\rm L}_{xxxz} &=& \frac{4 \beta_0}{1 + (\omega \tau)^2}\ , \label{L1}\\
\beta^{\rm L}_{xyyz} &=& - \frac{4 \beta_0}{ [1 + (\omega \tau)^2]^2 }\ ,\label{L2} \\
\beta^{\rm L}_{yxyz} &=& 4 \beta_0 \frac{2 + (\omega\tau)^2}{[1 + (\omega\tau)^2]^2}
\ ,\label{L3}\\
\beta^{\rm L}_{xyzy} &=& - \beta^{\rm L}_{yxzy} =
2\beta_0 \frac{1 -(\omega\tau)^2}{[1 + (\omega\tau)^2]^2}\ ,\label{L4}\\
\beta^{\rm C}_{yzz} &=& - 2\beta_0 \frac{ \omega \tau (1 -(\omega\tau)^2}
{[1 + (\omega\tau)^2]^2}\ ,\label{L5}\\
 \beta^{\rm C}_{xxy} &=& \beta^{\rm C}_{yyy} = - 2\beta_0
 \frac{\omega \tau}{[1 + (\omega\tau)^2]^2}\ ,\label{L6}
\end{eqnarray}
\end{subequations}
where $\beta_0 = A_s N_{2D} e^4 \tau^2/m^* c$.

It follows from the above equations that in the static limit
$\omega=0$ the magnetic-field induced corrections are given by
\begin{equation}
    \delta j_x= 2 A_s {e m^*}  \sigma_H (E_x^2-E_y^2), \quad
    \delta j_y= 4 A_s {e m^*}  \sigma_H E_x E_y
\end{equation}
at ${\bm B} \parallel z$, and
\begin{equation}
    \delta j_x= A_s {e m^*}  \sigma_H E_y E_z, \quad
    \delta j_y= - A_s {e m^*}  \sigma_H E_x E_z
\end{equation}
at ${\bm B} \parallel y$. Here $\sigma_H$ is the Hall conductivity
equal to $\omega_c \tau \sigma_0$, where $\omega_c$ is the
cyclotron frequency and $\sigma_0$ is the Drude conductivity at
$\omega=0$.

\section{Discussion and conclusion}

We have developed a semiclassical theory of nonlinear transport
and the photogalvanic effects in non-centrosymmetric media. It has
been shown that the terms in semiclassical kinetic equations
describing the electron motion associated with the Berry curvature
and elementary shifts in the real space contribute to the linear
and circular photogalvanic effects as well as to the quadratic I-V characteristics
(as a static limit of the LPGE). Previously Bloch et
al.\cite{Bloch1,Bloch2} studied the skew scattering mechanism of
the LPGE.
However, since it makes no
contribution to the CPGE the Berry-curvature and shift related
mechanisms certainly dominate the current sensitive to the
chirality of the electromagnetic wave.

It should be noted that we have considered linear-in-electric-field corrections to the collision integral due to the quantum coordinate shifts, see the term $- e {\bm E}(t) \cdot {\bm r}_{{\bm p}'{\bm p}}$ in the $\delta$-function of Eq.~\eqref{w2}. A comparable contribution
to the tensors $\bm \chi$ and $\bm \gamma$ comes from linear-in-$\bm E$ corrections to the squared matrix element $\left< \left\vert V_{{\bm p}'{\bm p}} \right\vert^2  \right>$ allowed in non-centrosymmetric systems.~\cite{ST}

The following estimations summarize different contributions and allow one to compare their
relative role in the formation of a \textit{dc} current as a quadratic
response to an \textit{ac} electric field.
The skew scattering contribution can be estimated as
$$
j_{\rm sk} \sim e \, N_{d} \, \bar{v} \, \xi^{({\rm imp})}_{\rm sk}  \frac{f_2^{(2)}}{f_0}
$$
or
\begin{equation} \label{chisc}
\chi_{\rm sk} \sim \xi_{\rm sk}^{({\rm imp})} {\rm Re}(\sigma_\omega)\frac{e l}{\bar{\varepsilon}} \:,
\end{equation}
where  $\bar{v}$ and $\bar{\varepsilon}$ are
the characteristic electron velocity and energy ($\bar{\varepsilon} \tau \gg \hbar$), $l=\bar{v} \tau$  is
the mean free path, and the
dimensionless parameter of the scattering asymmetry is given by,
see Eq.~\eqref{skew},
\begin{equation}
\xi_{\rm sk}^{({\rm imp})} =
\tau \, \overline{ {v_{{\bm p}\lambda} \over |{\bm v}_{\bm p}|} \: \Phi_{\bm p} },
\quad
\Phi_{\bm p}=I^{\rm (el,a)}_{\bm p}\left\{{v_{{\bm p}\mu} v_{{\bm p}\nu} \over |{\bm v}_{\bm p}|^2} \right\}.
\end{equation}
For static defects with a dipole moment ${\bm d}$ the asymmetry
parameter is estimated\cite{SturmanFridkin} as $d/(e a_B)$, where
$a_B$ is the Bohr radius $\ae \hbar^2/(m^* e^2)$, $\ae$ is the
static dielectric constant. If the dipole moment $d$ is anomalously small
one can take into account in the second term of the right-hand
side of Eq.~\eqref{skew} a combined scattering simultaneously
involving a defect and an acoustic phonon. The result is formally
obtained by the replacement of $\xi_{\rm sk}^{({\rm imp})}$ in
Eq.~\eqref{chisc} by $\hbar \xi_{\rm sk}^{({\rm ph})}/(\tau_{\rm ph} \bar{\varepsilon})$, where $\tau_{\rm ph}$ is the
electron momentum relaxation time due to scattering by acoustic
phonons, $\xi_{\rm sk}^{({\rm ph})}$ is the dimensionless
asymmetry parameter of the electron-phonon interaction including
the deformation-potential and piezoelectric mechanisms, see, e.g.,
Ref.~\onlinecite{GaAs}.

The Berry-curvature and shift-related contributions to LPGE, $\chi_{\rm B}$ and $\chi_{\rm sh}$, are given by the constants $A$ and $\eta$, respectively, see Eqs.~\eqref{chigammacxy2} and~\eqref{velC_s}.
In semiconducting pyroelectric materials one can use a general estimation $A \sim \eta \sim
\xi_{\rm piro} a_0 /(m^* E_g)$, where $a_0$ is lattice constant
and $\xi_{\rm piro}$ is the dimensionless asymmetry parameter.
As a result, we have
\begin{equation} \label{BerryShift}
\chi_{\rm B} \sim \chi_{\rm sh} \sim \xi_{\rm piro} {\rm Re}(\sigma_\omega) {e a_0 \over E_g}.
\end{equation}
For
a wurzite-type semiconductor one has from Eq.~\eqref{A_polar}
$\xi_{\rm piro} a_0 \sim \hbar Q / \Delta_c\:.$

It follows from Eqs.~(\ref{chisc}) and (\ref{BerryShift}) that,
for comparable asymmetry parameters $\xi_{\rm sk}^{({\rm imp})}$
and $\xi_{\rm piro}$, the skew-scattering contribution to the LPGE
prevails over the Berry-curvature and shift related contributions.

On the other hand, the
Boltzmann kinetic equation yields no contribution to the CPGE even if skew scattering is included: $\gamma_{\rm sk}=0$.
The circular nonlinear current appears only if ${\bm F}_{\bm p}$ and/or
${\bm r}_{{\bm p}'{\bm p}}$ are nonzero in which case one has
\[
	\gamma_{\rm B} \sim A e m^* {\rm Im}(\sigma_{\omega})\:, \quad
	\gamma_{\rm sh} \sim \eta e m^* {\rm Im}(\sigma_{\omega})\:.
\]
In the static limit both values $\gamma_{\rm B}$ and $\gamma_{\rm sh}$ as
expected vanish. At finite frequences, however, the Berry-curvature and shift-related effects provide
the main contribution to the CPGE.

\section*{Acknowledgements}

Work of L.~E.~G. and E.~L.~I. was  supported by the RFBR,
``Dynasty'' Foundation --- ICFPM, President grant for young
scientists, and DFG Merkator professorship. B.~S. and E.~D. were  supported by the NSF
grant DMR-0704151.

\end{document}